\documentclass[10pt,pre,reprint,showkeys,aps,fleqn]{revtex4-1}
\usepackage{amsmath}
\usepackage{amssymb}
\usepackage{amsfonts}
\usepackage{bm}
\usepackage{graphicx}
\usepackage{amsthm}
\usepackage{booktabs}
\usepackage{hyperref}

\usepackage{algcompatible}
\usepackage{newfloat}

\DeclareFloatingEnvironment[
    fileext=loa,
    listname=List of Algorithms,
    name=Algorithm,
    placement=tbhp,
]{algorithm}

\newcommand{\SO}[1]{SO(#1)}

\newcommand{\qmax}{q_\text{max}}

\newcommand{\defeq}{\overset{\Delta}{=}}
\newcommand{\ip}[1]{\left< #1 \right>} 
\newcommand{\nside}{n_{\text{side}}}

\newcommand{\npix}{n_{\text{pix}}}
\newcommand{\ie}{\emph{i.e.}\:}

\newcommand{\etal}{\emph{et al.\:}}

\newcommand{\Llim}{L^{\text{lim}}}

\renewcommand{\phi}{\varphi}

\begin{document}

\title{Expansion-maximization-compression algorithm with spherical harmonics for single particle imaging with X-ray lasers}
\date{\today}
\author{Julien Flamant}
\affiliation{Univ. Lille, CNRS, Centrale Lille, UMR 9189 - CRIStAL - Centre de Recherche en Informatique Signal et Automatique de Lille, 59000 Lille, France}
\author{Nicolas Le Bihan}
\affiliation{CNRS/GIPSA-Lab, 11 Rue des mathématiques, Domaine Universitaire, BP 46, 38402 Saint Martin d'Hères cedex, France}
\author{Andrew V. Martin}
\affiliation{ARC Centre of Excellence for Advanced Molecular Imaging, School of Physics, The University of Melbourne, Victoria, 3010, Australia.}
\author{Jonathan H. Manton}
\affiliation{Department of Electrical and Electronic Engineering The University of Melbourne, Victoria 3010, Australia}

\begin{abstract}
In 3D single particle imaging with X-ray free-electron lasers, particle orientation is not recorded during measurement but is instead recovered as a necessary step in the reconstruction of a 3D image from the diffraction data. Here we use harmonic analysis on the sphere to cleanly separate the angular and radial degrees of freedom of this problem, providing new opportunities to efficiently use data and computational resources. We develop the Expansion-Maximization-Compression algorithm into a shell-by-shell approach and implement an angular bandwidth limit that can be gradually raised during the reconstruction. We study the minimum number of patterns and minimum rotation sampling required for a desired angular and radial resolution. These extensions provide new avenues to improve computational efficiency and speed of convergence, which are critically important considering the very large datasets expected from experiment.

\end{abstract}
\maketitle

\section{Introduction}

	\subsection{Single-particle imaging with X-ray lasers}
	3D single-particle imaging with X-ray free-electron lasers (XFELs) is being actively pursued for its potential biological applications, which notably include determining the structures of single molecules \cite{Aquila2015}. It is also one of the most challenging goals in X-ray laser science, because each diffraction measurement provides a very weak signal and is destructive. As few as $5\times 10^{-2}$ photons per Shannon-Nyquist pixel (at high scattering angles) are expected for a 500 kDa protein \cite{fung2009structure}. To overcome low signal-to-noise, the experiment involves a large number of diffraction measurements (10$^4$--10$^6$) and each measurement must be of a new copy of the particle, which is assumed to have an identical structure. Such datasets are achievable because XFELs have high repetition rates ($>$ 100 Hz \cite{LCLS}) and serial injection technology has been developed to continuously deliver fresh sample into the beam path \cite{DePonte2008,Bogan2010b}. A consequence of this experimental approach is that particle orientation is not measured and must be determined by analyzing the diffraction data \cite{huldt2003diffraction}. In practice, a single orientation is not associated with each diffraction pattern, but rather the low signal per pattern is more effectively handled by treating the data globally to directly obtain a 3D diffraction intensity volume \cite{fung2009structure,loh2009reconstruction}. This intensity volume in then given as input for a phase retrieval algorithm that produces an image of the particle \cite{ekeberg2015machine}.

	\subsection{Intensity reconstruction strategies}
		 In order to assemble the two-dimensional noisy diffraction patterns into a consistent three-dimensional intensity function, numerous algorithms have been proposed to date. Early work from Huldt \etal \cite{huldt2003diffraction} and Bortel \etal \cite{Bortel2009226} classified patterns into classes, which were averaged to improve signal-to-noise prior to orienting the classes using common arcs of intersection. However, the classification is not suffiently accurate to handle the low signals expected in experiment \cite{shneerson2008crystallography}. This approach has since been superseded by a variety of methods that treat the data globally to overcome the issue of noise.


		Fung \etal \cite{fung2009structure} proposed an algorithm based on a manifold embedding technique. This approach is based on a Generative Topographic Mapping (GTM), where each pattern is considered as a vector of the $N$-dimensional space of intensities, with $N$ the number of pixels on the detector. Since a continuous rotation of the sample implies a continuous variation of the diffraction intensities, the images obtained should span a three-dimensional manifold embedded in $N$-dimensional space. The manifold is generated from a large number of diffraction patterns, and averaging out the closest diffraction patterns leads to the expected smooth manifold. However, a large number of diffraction patterns may be required to obtain a sufficiently smooth manifold \cite{barty2013molecular}. We note that this approach has been developed further recently in \cite{Hosseinizadeh20130326}. More recently, diffusion map techniques have been developed to compute low-dimensional manifolds from XFEL diffraction data \cite{Giannakis2012}. In practice, these techniques can generate more than three significant dimensions revealing other experimental variables such as changing beam conditions or sample heterogeneity. The advantage here is that the generation of the manifold is not biased by human assumptions. The challenge is the interpreting the manifold and identifying the correct relationship between the three degrees of freedom on the manifold and the rotation group. One way to explicitly relate the data-space to rotations is via mapping geodesics \cite{Kassemeyer2013}.
		
		The first attempt to introduce geometrical constraints in the processing of diffraction patterns was taken by Saldin \etal \cite{saldin2009structure}, where the three-dimensional intensity function was expanded on the spherical harmonic basis. Their approach was based the cross-correlation of the diffraction patterns and exploits the orthogonality of spherical harmonics to obtain a decomposition of the cross-correlation function. It is interesting to note that this method has been tested experimentaly \cite{starodub2012single} on large dimers with known structure. For objects of known rotational symmetry, the correlation function can be converted into the 3D single-particle Fourier intensity, which can then be converted to an image of the particle via phasing. For some time, it was known how to extend this method to an asymmetric 3D object, but recently an algorithm has been proposed to phase directly from the correlation function \cite{Donatelli18082015}. The correlation function can also be used with molecular replacement techniques \cite{saldin2009structure}. It has the added advantage that multiple particles can be illuminated per measurement, which not many other algorithms can handle.

		In Ref. \cite{loh2009reconstruction}, Loh and Elser have introduced the Expansion-Maximization-Compression (EMC) algorithm which relies on an Expectation-Maximization (EM) technique. As a Bayesian method, it aims to use known information about the noise statistics to handle very low signal-to-noise levels. The algorithm iteratively maximizes the likelihood of the reconstructed intensity given the set of diffraction patterns. The algorithm does not assign a single molecular orientation to each pattern, but rather it estimates the probability for a diffraction pattern to be associated with a certain orientation. Recently, this algorithm has proven its feasibility with the reconstruction of Mimivirus from experimental data collected at LCLS \cite{ekeberg2015three}. The EMC algorithm is studied in detail in this work, with the introduction of several improvements. Recently, Walczak \etal \cite{walczak2014bayesian} developed the Bayesian approach further using seed structural models to speed up convergence and to discriminate between different conformations. Finally, we note that Tegze and Bortel \cite{Tegze201241} have proposed a simplified version of the EMC algorithm, where the diffraction pattern orientation is fully assigned through the intensity best fit, rather than in a probalistic way.

	A drawback of the EMC algorithm is that it is computationally expensive to implement on large datasets that are expected in experiment. The EMC algorithm stores a matrix that grows proportionally with the number of diffraction patterns and must be recalculated at every iteration. This is not a trivial step if there are 10$^5$ -- 10$^6$ measured diffraction patterns. Hence, there is still scope to develop the EMC algorithm further to enable parallelization and efficient computation. Another related issue is analyzing smaller datasets at lower resolution, which would be very valuable during an experiment to provide feedback on data collection. Algorithms that are readily scaled with the number of patterns and resolution are also valuable.
			
	With these considerations in mind, we study here the application of spherical harmonic analysis to the EMC algorithm. Our goal is to produce an EMC algorithm that can be scaled in terms of angular and radial resolution. This facilitates starting with lower resolution reconstruction with less patterns, higher signal and less computational time, before proceeding to higher resolutions. By implementing the algorithm shell by shell, we also enable it to be easily parallelized. After reviewing the relevant properties of spherical harmonics in Section \ref{sec:geom}, we formulate the EMC algorithm for a single q-shell in Section \ref{sec:EMC} and impose an angular bandwidth limit. Section \ref{sec:shell} describes how reconstructions on neighbouring shells can be aligned via correlations, and finally numerical tests are provided in Section \ref{sec:numeric}.
	
While our work improves upon the computational requirements for EMC, it does not match the computational advantages of correlation methods based on spherical harmonics  that do not require more memory as the number of patterns increases\cite{saldin2009structure}. As described above, however, correlation-based methods have other issues for particles without rotational symmetry, while EMC has potential advantages for analyzing low signal data, which contributes to its popularity. Explorations of how to retain the advantages of EMC at lower computational expense are thus valuable for the development of XFEL single particle imaging.

\section{The geometry of intensity functions}
\label{sec:geom}
	
	The intensity function of a biomolecule $I(\bm q)$ is linked to $F(\bm q)$ the Fourier transform of its 3D electron density $\rho^{\text{(mol)}}(\bm r)$, with $\bm r$ and $\bm q$ respectively the real and reciprocal space coordinates. During XFEL experiment, each diffraction pattern is obtained from a randomly rotated copy of the biomolecule. A rotation of the molecule in the real space corresponds to the same rotation around the origin of the Ewald sphere in the dual space. The accumulation of thousands of diffraction patterns from rotated molecules can cover a spherical volume in the reciprocal space, in order the reconstruct of the 3D intensity of the biomolecule. 




	With the notation previously introduced, the Fourier transform or molecular transform $F$ of the three-dimensional electron density of the molecule reads,
	\begin{equation}
	F(\bm q) \defeq \int \rho^{\text{(mol)}}(\bm r)\exp\left(\mathrm{i}\bm q\cdot\bm r\right)\mathrm{d}^3\bm r \;.
	\end{equation}
	The intensity function $I$ is defined as the square magnitude of the Fourier transform of the electron density, up to a constant normalizing factor $I_0$ depending on the experimental conditions \cite{chapman2006high}. Thus we have the following
	\begin{equation}
	I(\bm q) = \left\vert F(\bm q) \right\vert^2 I_0.
	\end{equation}

	In what follows, we will be using spherical coordinates for $\bm q$, meaning that its parametrization will be
	\begin{equation}
	\bm q = (q\sin\theta\cos\phi, q\sin\theta\sin\phi,q\cos\theta)^T,
	\end{equation}
	where $q = \Vert \bm q \Vert$ is the radial coordinate, and $(\theta,\phi) \in [0,\pi)\times[0,2\pi)$ are angular coordinates, where the polar angle is relative to the beam axis. In the following we make use of the shorthand notation $\bm q = (q,\bm \Omega)$, where $\bm \Omega$ is an unit vector pointing towards angular coordinates $(\theta,\phi)$.

	In order to take advantage of the symmetries of the intensity function $I$, we propose to consider it as a set of concentric shells of increasing radii. On such a shell of radius $q_s$, where $s$ stands for the shell number, we define the intensity  $I^s(\bm \Omega) = I(q_s, \bm \Omega)$. In the reconstruction process, the three-dimensional intensity function will be recovered by assembling the reconstructed intensities over spherical shells.

	The shell-by-shell reconstruction can be made efficient using the fact that each $I^s$ can be decomposed on a spherical harmonic basis. Such spherical basis is presented is Section  \ref{subsec:HarmdcompI}. Before introducing this decomposition, we first present the measurement model.







	\subsection{Equivalent single shell measurement model}\label{section:shell_measurement}
		

		
\begin{figure*}
		\includegraphics[width=\textwidth]{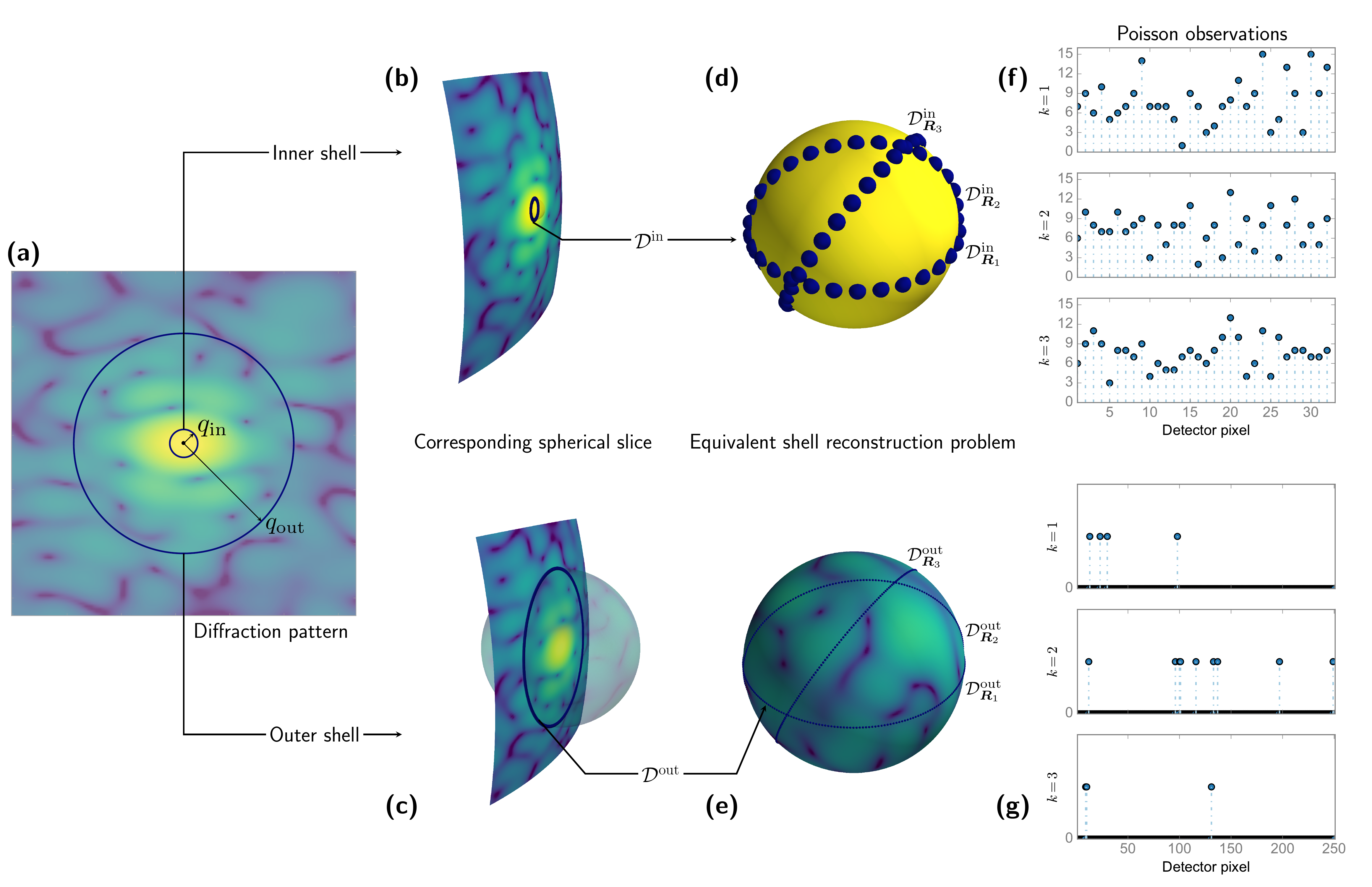}
		\caption{Shell-by-shell processing procedure. \textbf{\textsf{(a)}} Typical diffraction pattern of the small virus STNV (PDB entry: 2BUK). Circles represent two configurations, corresponding to inner and outer shell respectively. \textbf{\textsf{(b)}} and \textbf{\textsf{(c)}} Corresponding Ewald sphere slice through the corresponding shell, whose intersection is given by a circle. \textbf{\textsf{(d)}} and \textbf{\textsf{(e)}} Equivalent single shell measurement process. Diffraction patterns on the shell as randomly rotated copies of the detector sampling points. \textbf{\textsf{(f)}} and \textbf{\textsf{(g)}} Corresponding Poisson realizations of the sampling points depicted in \textbf{\textsf{(d)}} and \textbf{\textsf{(e)}}.}\label{fig:shellProcessingMethod}
		\end{figure*}

		In order to perform shell-by-shell reconstruction of the intensity function, we introduce a measurement model which relates diffraction data to the corresponding part of the intensity on each shell.	
		In this model, as shown in figure \ref{fig:shellProcessingMethod}, the corresponding \emph{diffraction patterns} on a spherical shell $s$ are mono-dimensional, {\em i.e.} circles on this spherical shell. Formally, we denote by $\mathcal{D}^s$ the \emph{detector} on the shell, that is the reference sampling points. $\mathcal{D}^s$ is defined as the set of $N_s$ \emph{pixels} coordinates on the spherical shell $I^s(\bm \Omega)$ and given by
		\begin{equation}\label{eq:detector_definition_shell}
		\mathcal{D}^s \defeq \left\lbrace  \bm q_i  = (q_s,\bm \Omega_i) \middle\vert i = 1,2,\ldots,N_s\right\rbrace,
		\end{equation}
		where $\bm \Omega_i$ is given in angular coordinates by $(\phi_i, \theta(q_s)) \in [0,2\pi)\times [0,\pi]$, and $q_s$ is the radius of the spherical shell. The dependence of $\theta$ on the radius $q_s$ is a consequence of the scattering geometry. For completeness we recall that
		\begin{equation}
		\theta(q_s) = \frac{\pi}{2} - \arcsin \frac{q_s}{\qmax},
		\end{equation}
		where $\qmax$ is defined in terms of the wavelength $\lambda$, $\qmax = 4\pi/\lambda$.

		On the detector, the number of pixels contributing to a shell, \ie the number of pixels on a circle on the diffraction pattern increases with the radius $q_s$. The number of points $N_s$ on the detector are scaled such that:
		\begin{equation}
		N_s = \left\lceil \frac{2\pi q_s}{\Delta q }\right\rceil, \quad \phi_i = \frac{2\pi}{N_s}i,\: i \in \left\lbrace 0,1,\ldots, N-1\right\rbrace,
		\end{equation}
		where $\lceil \cdot\rceil$ denotes the ceiling function, and $\Delta q$ is the reciprocal space pixel size.

		Finally, each measurement $k$, or diffraction pattern on a shell consists in a set of $N_s$ samples $y_{ik}$,  $i=1,\ldots, N_s$ given by a Poisson model \cite{elser2009noise} (due to low-photon counts expected in single-particle experiments)
		\begin{equation}
		y_{ik} = \mathrm{Pois}\left(I^s(\bm R_k\cdot \bm \Omega_i)\right), \: \bm R_k \sim \mathsf{U}(\SO{3})
		\end{equation}
		where the random rotation $\bm R_k$ is drawn from the uniform distribution on the rotation group $\SO{3}$.
	\subsection{Harmonic decomposition of the shell intensity}
	\label{subsec:HarmdcompI}
		 The intensity on a given shell $s$, {\em i.e.} $I^s(\bm \Omega)$, can be expanded on the spherical harmonics basis (see appendix \ref{appendix_A} for details) as follows,
		\begin{equation}\label{eq:SHIntensityModelGeneric}
		I^s(\bm \Omega) = \sum_{\ell=0}^\infty\sum_{m=-\ell}^\ell c_\ell^m(q_s)Y_\ell^m(\bm \Omega) \;,
		\end{equation}
		where the coefficients $c_\ell^m(q_s)$ are the spherical harmonic coefficients of the intensity function on the shell $s$. It turns out that the spherical harmonic representation provide an efficient way to deal with the symmetries of the intensity functions. In the following, we exploit the physical properties of the intensity function to simplify further the expansion above. 

		First, intensity functions are by definition real-valued functions. As a consequence, its spherical harmonic coefficients exhibits the well-known conjugation property of spherical harmonics
		\begin{equation}
		\text{for all } \ell \geq 0, \: c_\ell^{-m} = (-1)^m\overline{c_\ell^m}, \: m \in \left\lbrace 0,1,\ldots, \ell\right\rbrace
		\end{equation}
		Moreover, another interesting property is the centro-symmetry of intensity functions, also known as the Friedel property. This symmetry property reads $I^s(\bm \Omega) = I^s(-\bm \Omega)$ and it is straightforward to see that with this constraint every coefficient of odd degree $\ell$ is set to zero,
		\begin{equation}\label{eq:FriedelSymmetryCoefficients}
		\text{for all } p \geq 0,\: c_{2p+1}^m = 0, \: m \in \left\lbrace 0,1,\ldots, 2p+1\right\rbrace
		\end{equation}
		Finally, the finite size nature of the molecule in real space implies that the spectral representation on each shell is effectively bandlimited, meaning that the coefficients in the spherical harmonics expansion are non-zero up to a maximum degree $\Llim_s$
		\begin{equation}
		I^s(\bm \Omega) = \sum_{\ell = 0}^{\Llim_s-1}c_\ell^m(q_s)Y_\ell^m(\bm \Omega)
		\end{equation}
		where $\Llim_s$ is the bandlimit on the shell $s$, and where the sum has been restricted to even values of the degree $\ell$. Since Friedel symmetry imposes condition (\ref{eq:FriedelSymmetryCoefficients}) on the coefficients, the value of $\Llim_s$ is an odd number. 

		It is interesting to note that rotational symmetries of the particle will further reduce the number of independent coefficients. This makes it easier to implement particle symmetry than for the Cartesian sample of the intensity function used in the original EMC algorithm and may improve performance in the case of highly symmetric particles. 

		\begin{figure}
		\centering
		\includegraphics[width = 0.5\textwidth]{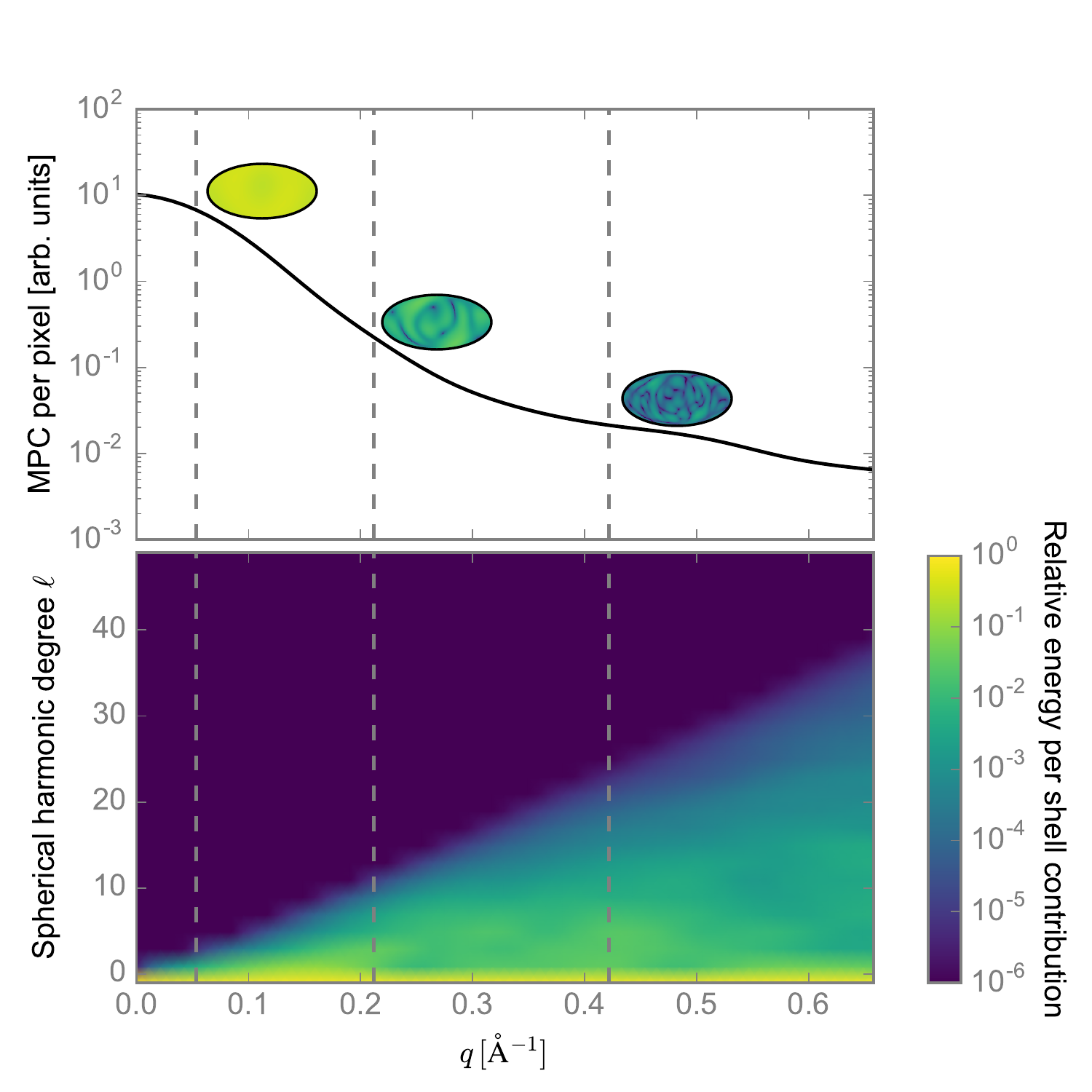}
		\caption{(Top) Mean Photon Count (MPC) per pixel as a function of the reciprocal space magnitude $q$. The signal becomes weaker as $q$ increases. Vertical lines show three examples of shell intensity distributions, refered as inner, middle and outer shell therein. (Bottom) Relative energy per shell distribution in the spherical harmonic domain, for increasing values of $q$. The distribution spreads with increasing values of $q$.}\label{fig:theoreticalEnergyDistribution}
		\end{figure}

\section{Spherical EMC algorithm}
\label{sec:EMC}
	We develop in this section extension of the original EMC algorithm  \cite{loh2009reconstruction} to the spherical harmonic basis \footnote{A working implementation of the algorithms described in this paper is avalaible at \url{http://github.com/jflamant/sphericalEMC}.}.Precisely, we define an update rule in terms of the spherical harmonic coefficients of the shell intensity, and scale efficiently the computation time in terms of the bandlimit $L$. 

	In what follows, all quantities with superscript $^{(n)}$ denotes the corresponding values at the $n$-th iteration of the EM algorithm.

	\subsection{A general Expectation-Maximization framework}
	We start this section by recalling the framework of EM algorithms.
	In general, EM algorithms provide an efficient way to deal with missing data \cite{dempster1977maximum}. In our spherical setting, we design an EM algorithm which updates the spherical harmonics coefficients at each iteration, namely we write the update rule 
	\begin{equation}
	\hat{\bm c}^{(n)} \rightarrow \hat{\bm c}^{(n+1)},
	\end{equation}
	where $n$ denotes the iteration index. Note that we distinguish the estimated coefficients $\hat{\bm c}$ from the theoretical ones $\bm c$ to avoid confusion.
	At each iteration of the EM algorithm, the likelihood of the estimated parameter $\hat{\bm c}$ given the set of measurements $\bm y_1, \ldots, \bm y_k$ is increased. However, EM-algorithms do not work directly with the likelihood function, but rather with the \emph{intermediate quantity} $\mathcal{Q}(\hat{\bm c} \vert \hat{\bm c}^{(n)})$, whose value is determined in the expectation step (e-step) \footnote{We use the non-capital ``e'' letter for the term ``expectation'', whereas the capital ``E'' is reserved for ``Expansion''.} of the algorithm. For practical purposes, we first introduce the intermediate quantity per pattern $\mathcal{Q}_k(\hat{\bm c} \vert \hat{\bm c}^{(n)})$, which reads
	\begin{equation}\label{likelihood_pattern}
	\mathcal{Q}_k(\hat{\bm c }\vert \hat{\bm c}^{(n)})  = \int_{\SO{3}}\log\left[ p(\bm y_k,\bm R\vert\bm \hat{\bm c})\right]p(\bm R\vert \bm y_k, \hat{\bm c}^{(n)})\mathrm{d}\mu(\bm R)
	\end{equation}
	where $\mathrm{d}\mu(\bm R)$ denotes the (bi)-invariant Haar measure on $\SO{3}$. Since the acquisitions $\bm y_1, \ldots, \bm y_K$ are independent from each other, we can express the total intermediate quantity as the sum
	\begin{equation}
	\textbf{(e-step)}:\quad 
	\mathcal{Q}(\hat{\bm c} \vert \hat{\bm c}^{(n)}) = \frac{1}{K} \sum_{k=1}^K \mathcal{Q}_k(\hat{\bm c}\vert \hat{\bm c}^{(n)}),
	\end{equation}
	where the choice of the normalization made here is arbitrary.

	The maximization step can now be defined as the maximization of the (total) intermediate quantity, such that it reads
	\begin{equation}
	\textbf{(M-step)}: \quad \hat{\bm c}^{(n+1)} = \underset{\hat{\bm c}}{\arg \max} \:\mathcal{Q}(\hat{\bm c }\vert \hat{\bm c}^{(n)}) 
	\end{equation}
	We note here that there are potentialy several values of $\hat{\bm c}$ that maximize the intermediate quantity; therefore uniqueness is not guaranteed. This is a general property of expectation-maximization algorithms.

\subsection{Proposed Spherical EMC algorithm (SEMC)}
	\subsubsection{EM by tomographic grid update}
		Considering the EM algorithm expressed directly with the spherical harmonic coefficients is a complicated task, which leads to intractable maximizations procedures. Rather than solving the problem directly in the spherical harmonic domain, we extend the original approach of Loh and Elser \cite{loh2009reconstruction} to a spherical framework.

		Let us introduce a spherical intensity model, denoted by $W$. This spherical intensity model $W$ can be expressed using a spherical harmonic expansion similar to (\ref{eq:SHIntensityModelGeneric}). The corresponding intermediate quantity per pattern reads
		\begin{equation}
		\begin{split}
		\mathcal{Q}_k(W\vert W^{(n)}) = \int_{\SO{3}}&\log\left[p(\bm y_k,\bm R\vert W_{\bm R})\right]\\ & \times p(\bm R\vert \bm y_k, W^{(n)}_{\bm R})\mathrm{d}\mu(\bm R)
		\end{split}
		\end{equation}
		where $W_{\bm R}$ is a shorthand notation for the set $\left\lbrace W_{i,\bm R} = W(\bm R\cdot q_i), \: i=1,\ldots,N\right\rbrace$. Recall that the $\bm q_i$ denote the reciprocal space coordinates of the detector. 

		Now, let us give details on the different quantities above. The conditional probabilities read
		\begin{align}
		p(\bm y_k, \bm R\vert W_{\bm R}) &= p(\bm y_k\vert \bm R,W_{\bm R})p(\bm R\vert W_{\bm R})\\
		p(\bm R\vert \bm y_k, W^{(n)}_{\bm R}) &= \frac{p(\bm y_k\vert \bm R,W^{(n)}_{\bm R})p(\bm R\vert W_{\bm R}^{(n)})}{p(\bm y_k \vert  W_{\bm R}^{(n)})}
		\end{align}
		The first equality is simply the definition of the joint probability, whereas the second equality is obtained applying Bayes' rule. We note that those quantities involve the prior distribution on the orientations, {\em i.e.} $p(\bm R\vert W_{\bm R}) = p(\bm R\vert W_{\bm R}^{(n)}) = p(\bm R)$ since the model $W$ is assumed deterministic. We assume there is no preferential orientation of the particle and that the distribution of orientations is uniform, which means with respect to the bi-invariant Haar measure on $\SO{3}$ that
		\begin{equation}
		p(\bm R) = 1, \: \forall\: \bm R \in \SO{3}.
		\end{equation}
		Moreover, the Poisson assumption on the photon count and the independence of each pixel yields to 
		\begin{equation}
		p(\bm y_{k}\vert \bm R, W_{\bm R}) = \prod_{i=1}^N\frac{(W_{i,\bm R})^{y_{ik}}}{y_{ik}!}\exp(-W_{i,\bm R})
		\end{equation}
		which gives the expression of the intermediate quantity per pattern
		\begin{widetext}
		\begin{equation}
		\mathcal{Q}_k(W\vert W^{(n)}) = Z_1 + Z_2^{-1}\int_{\SO{3}}\mathrm{d}\mu(\bm R)\left[ \sum_{i=1}^N y_{ik}\log W_{i,\bm R} - W_{i,\bm R}\right]\prod_{i=1}^N(W_{i,\bm R}^{(n)})^{y_{ik}}\exp(-W_{i,\bm R}^{(n)})
		\end{equation}
		\end{widetext}
		where $Z_1$ is a constant only depending on the data $\bm y_k$ (which is arbitrarly removed hereafter), and $Z_2$ is a normalization constant given by
		\begin{equation}
		Z_2 = \int_{\SO{3}} \mathrm{d}\mu(\bm R)\prod_{i=1}^N(W_{i,\bm R}^{(n)})^{y_{ik}}\exp(-W_{i,\bm R}^{(n)}).
		\end{equation}

		Now, we face the following problem of computing the integrals over the rotation group $\SO{3}$. Unfortunately there is no exact way to do so due to the lack of quadrature formula for probability functions on the rotation group $\SO{3}$. Nevertheless, the integrals above can be approximated via a discrete sum over elements of the rotation group. More precisely, if we suppose that we have some sampling set $\mathcal{X} \subset \SO{3}$ of size $M$ and whose elements are indexed by $j$, the intermediate quantity reads
		\begin{widetext}
		\begin{equation}
		\mathcal{Q}_k(W\vert W^{(n)}) \simeq \tilde{Z}_2^{-1}\sum_{j=1}^M w_j \left[ \sum_{i=1}^N y_{ik}\log W_{ij} - W_{ij}\right]\prod_{i=1}^N(W_{ij}^{(n)})^{y_{ik}}\exp(-W_{ij}^{(n)})
		\end{equation}
		\end{widetext}
		where $w_j$ are equivalent to quadrature weights such that $\sum_j w_j = 1$, $W_{ij} \defeq W_{i,\bm R_j}$, and where $\tilde{Z}_2$ is obtained following the same guidelines. Since the model $W$ is now evaluated at $M$ rotated versions of the original detector coordinates, we call it a \emph{tomographic model}. In a similar way, we define the \emph{tomographic grid} coordinates as $\bm q_{ij} = \bm R_j\cdot \bm q_i$.

		In short-form, we can rewrite the latter intermediate quantity as a function of the tomographic model
		\begin{equation}
		\mathcal{Q}_k(W\vert W^{(n)}) \simeq \sum_{j=1}^M P_{jk}\left(\sum_{i=1}^Ny_{ik}\log W_{ij}-W_{ij}\right)
		\end{equation}
		where $P_{jk}$ is given by
		\begin{equation}\label{eq:definition_proba_Pjk}
		P_{jk} = \frac{w_j\prod_{i=1}^N(W_{ij}^{(n)})^{y_{ik}}\exp(-W_{ij}^{(n)})}{\sum_{j=1}^Mw_jP_{jk}}
		\end{equation}
		Finally, the corresponding Expectation step of the EM algorithm is given by
		\begin{equation}\label{eq:intermediate_quantity_lohELSER}
		\mathcal{Q}(W\vert W^{(n)}) \simeq \sum_{k=1}^K\sum_{j=1}^M P_{jk}\left(\sum_{i=1}^Ny_{ik}\log W_{ij}-W_{ij}\right)
		\end{equation}
		The last expressions (\ref{eq:definition_proba_Pjk}) and (\ref{eq:intermediate_quantity_lohELSER}) are exactly the same as in Ref. \cite{loh2009reconstruction}, as expected. The expression of the intermediate quantity (\ref{eq:intermediate_quantity_lohELSER}) is rather simple and allows for a closed-form maximization procedure, that is
		\begin{equation}
		W^{(n+1)}_{ij} = \frac{\sum_{k=1}^KP_{jk}y_{ik}}{\sum_{k=1}P_{jk}}, \: i = 1,2, \ldots, N, j = 1,2, \ldots, M
		\end{equation}

		We note that in recent papers \cite{ekeberg2015three,ekeberg2015machine}, where the EMC algorithm is applied to experimental data, an update rule of the fluence in each diffraction pattern is also derived. Here we make the simplifying assumption that the fluence in each diffraction pattern is constant.

	\subsubsection{Spherical harmonics update through Expansion-Compression steps}

		By introducing a tomographic model $W$, the EM-procedure has been made easier. However, our goal is to obtain an algorithm which updates the spherical harmonic coefficients up to some degree $L-1$ for the considered shell, and this is done by adding two extra steps, known as Expansion and Compression. 

		First, we remark that the tomographic grid coordinates are treated independently: therefore there exists a lot of tuples $(i,j) \neq (i',j')$ such that the coordinates $\bm q_{ij}$, $\bm q_{i'j'}$ are really close. To enforce the consistency of the estimated model, we introduce a \emph{regular} spherical grid $\mathcal{G}_L$ and the corresponding regular model $W_{\mathcal{G}_L}$, whose nodes are given by $\bm q_p$. Note that the grid depends on the bandlimit $L$, allowing efficient scaling of the grid size. If the grid is well chosen and exhibits nice features, then the spherical harmonics coefficients $c_\ell^m$ up to some degree $L-1$ can be obtained using a fast implementation of the spherical harmonic transform. 

		The above description actually corresponds to the compression step (C-step), which can be schematically given by 
		\begin{equation}
		(\textbf{C-step}): W_{ij} \rightarrow W_{\mathcal{G}_L} \rightarrow \hat{\bm c},
		\end{equation}
		whereas the reverse operation is the expansion step (E-step)
		\begin{equation}
		(\textbf{E-step}):  \hat{\bm c} \rightarrow W_{\mathcal{G}_L} \rightarrow W_{ij}
		\end{equation}

	\subsubsection{Grid $\mathcal{G}$ and implementation of Expansion and Compression steps}
		We make use of the material introduced in appendix \ref{appendix_A} by choosing the HEALPix sampling scheme on the sphere for our grid $\mathcal{G}_L$. HEALPix grid on the sphere provides equal-area pixelization of the sphere, along with hierarchical resolution and numerous features \cite{gorski2005healpix}. The compression step is performed as follows. We first determine the tomographic points $\bm q_{ij}$ belonging to each HEALPix pixel, and then the intensity value on this pixel is given by Inverse Distance Weighting (IDW) between the respective $\bm q_{ij}$ and the pixel center $\bm q_p$,
		\begin{equation}
		W_{\mathcal{G}_L}(\bm q_p) = \frac{\sum_{\text{neighbors}}W_{ij}/{\Vert \bm q_{ij} - \bm q_p\Vert^2}}{\sum_{\text{neighbors}} 1/{\Vert \bm q_{ij} - \bm q_p\Vert^2} }.
		\end{equation}
		The coefficients $c_\ell^m(q_s)$ are then computed up to order $L-1$ as given by the sampling theorem (\ref{eq:appendix_sampling_theorem_Healpix}) using a fast Spherical Harmonic Transform (SHT). The Friedel symmetry is restored by canceling the coefficients for odd values of $\ell$. 

		The expansion step is done by the successive expansion of the coefficients $c_\ell^m(q_s)$ to obtain $W_{\mathcal{G}_L}$, then by computation of the tomographic intensities by interpolation on the sphere. Precisely the intensity $W_{\mathcal{G}_L}$ is obtained by inverse SHT, again implemented using the HEALPix package routines. The interpolation from the regular grid $\mathcal{G}$ to the tomographic grid is done by bilinear interpolation using the four nearest-neighbors on the regular grid. 
	
	\subsection{Implementation}
	Several issues need to be considered when using the proposed spherical EMC algorithm and the adaptive spherical EMC algorithm given in pseudocode algortihms \ref{alg1} and \ref{alg2} respectively. We discuss thereafter the critical points to control in order to ensure proper behaviour of the reconstruction algorithms. 
	\begin{algorithm}
	\caption{Spherical EMC algorithm}
 	\label{alg1}
	\begin{algorithmic}[1]
	\STATE{\textbf{Input}: Data $\bm y = (\bm y_1, \bm y_2, \ldots, \bm y_K)$, bandlimit $L$}
	\STATE{$\mathcal{G}_{ij} \leftarrow$ GenerateTomographicGrid($L$)}
	\STATE{$\mathcal{G}_L \leftarrow$ CreateHEALPixGrid($L$)}
	\STATE{$\hat{\bm c}^{(0)} \leftarrow $ RandomInit()}
	\STATE{$ n \leftarrow 0$}
	\WHILE{$\nabla \mathcal{L} > \eta$}
		\STATE{\emph{/*Expansion step*/}}
		\STATE{$W^{(n)}_{\mathcal{G}_L} \leftarrow $ invSHT($\hat{\bm c}^{(n)}$)}
		\STATE{$W^{(n)}_{ij} \leftarrow $ Regular2Tomo($W_{\mathcal{G}_L}^{(n)}$)}
		\STATE{\emph{/*Maximization step*/}}
		\STATE{$P_{jk} \leftarrow $ ComputeProbabilities($W_{ij}^{(n)}$, $\bm y$)}
		\STATE{$\mathcal{L}^{(n)} \rightarrow$ ComputeLikelihood($W_{ij}^{(n)}, \bm y, P_{jk}$)}
		\STATE{Compute $\nabla \mathcal{L}$}
		\STATE{$W_{ij}^{(n+1)} \leftarrow$ UpdateTomographicModel($P_{jk}$, $\bm y$)}
		\STATE{\emph{/*Compression step*/}}
		\STATE{$W^{(n+1)}_{\mathcal{G}_L} \leftarrow $ TomoToRegular($W_{ij}^{(n+1)})$}
		\STATE{$\hat{\bm c}^{(n+1)} \leftarrow$ SHT($W^{(n+1)}_{\mathcal{G}_L})$, $L$)}
		\STATE{$\hat{\bm c}^{(n+1)} \leftarrow $ FriedelSymmetry($\hat{\bm c}^{(n+1)}$)}
		\STATE{$n \leftarrow n+1$}
	\ENDWHILE
	\STATE{\textbf{Return} SH coefficients $\hat{\bm c}^{(n)}$}
	\end{algorithmic}
	\end{algorithm}

	\begin{algorithm}
	 \caption{Adaptive Spherical EMC algorithm}
 	\label{alg2}
	\begin{algorithmic}[1]
	\STATE{\textbf{Input}: Data $\bm y = (\bm y_1, \bm y_2, \ldots, \bm y_K)$, maximum bandlimit $L_\text{max}$}
	\FOR{$L \in \left\lbrace3,5, \ldots, L_\text{max}\right\rbrace$}
		\STATE{$\mathcal{G}_{ij} \leftarrow$ GenerateTomographicGrid($L$)}
		\STATE{$\mathcal{G}_L \leftarrow$ CreateHEALPixGrid($L$)}
		\IF{$L = 3$}
			\STATE{$\hat{\bm c}^{(0)} \leftarrow $ RandomInit()}
		\ELSE
			\STATE{$\hat{\bm c}^{(n)} \leftarrow$ InheritFromPreviousBandlimit()}
		\ENDIF
		\STATE{$\hat{\bm c}^{(n)} \leftarrow $ IterationsEMC($L$, $\bm y$)}
	\ENDFOR
	\STATE{\textbf{Return} SH coefficients $\hat{\bm c}^{(n)}$}
	\end{algorithmic}
	\end{algorithm}
	\subsubsection{Initialization}
	In general, initialization of EM-like algorithms is critical. Indeed, convergence is often guaranteed only to a local maximum of the likelihood. As a consequence, a bad initialization of the algorithm may lead to local convergence to an undesired non-global minimum. In order to minimize such an effect, an initialization procedure based on the bandlimit adaptive scheme proposed throughout this paper can be used. Namely, since the algorithm estimates bandlimited intensity functions with increasing degree, it seems reasonable to initialize the algorithm with the previous estimated model, at bandlimit $L-2$. At the first stage of the algorithm, corresponding to bandlimit $L = 3$, this is done the following way. First the tomographic model is initialized with the Mean Photon Count (MPC) of the diffraction patterns on this shell. Note that the MPC is defined here as an average over the detector pixels, not in terms of Shannon-Nyquist pixels. Then, to avoid the algorithm stopping prematurely, we slightly pertubate the MPC. The initialized tomographic model then reads
	\begin{equation}
		W_{ij}^{(0)} = \text{MPC} + \varepsilon u_{ij}, 
	\end{equation}
	with
	\begin{equation}
	\text{MPC} = \frac{1}{N_sK}\sum_{i,k} y_{ik}
	\end{equation}
	and where the $u_{ij}$'s are drawn uniformly on the interval $[-\text{MPC}, \text{MPC}]$. Here $\varepsilon$ is a randomization parameter, $0 < \varepsilon < 1$. Finally, performing one compression step leads to the estimate of initial spherical harmonic coefficients $\hat{\bm c}^{(0)}$, which completes the initialization step.
	
	\subsubsection{Convergence assessment}
	Another important point issue is the convergence assessment of the algorithm. In the general EM algorithm setting, this is often done by monitoring the likelihood value, whose increments are given by the so-called \emph{fundamental inequality of EM} \cite{dempster1977maximum,cappe2006inference}
	\begin{equation}
	\mathcal{L}(W') - \mathcal{L}(W) \geq \mathcal{Q}(W'\vert W) - \mathcal{Q}(W\vert W).
	\end{equation}
	The EM algorithm then stops when likelihood increments become smaller than a certain threshold value. In the EMC setting we added two extras steps, expansion and compression respectively, to make the Expectation-Maximization procedure tractable. This yields to a loss of important properties of the EM algorithm, namely the likelihood $\mathcal{L}(W)$ is not forced to increment at each iteration, and the likelihood value is not directly available. 

	To obtain the likelihood value, it is thus necessary to proceed as follows. First recall that we have only defined the intermediate quantity $\mathcal{Q}(W \vert W^{(n)})$ for now. The likelihood is obtained the following way
	\begin{equation}
	\mathcal{L}(W) = \mathcal{Q}(W\vert W^{(n)}) + \mathcal{H}(W\vert W^{(n)})
	\end{equation}
	where the \emph{entropy} $\mathcal{H}(W\vert W^{(n)})$ is defined like in \cite{cappe2006inference} by
	\begin{equation}
	\begin{split}
		\mathcal{H}(W\vert W^{(n)}) = -\frac{1}{K}\sum_{k=1}^K &\int_{\SO{3}}\log\left[p(\bm R\vert \bm y_k, W_{\bm R})\right]\\&\times p(\bm R\vert \bm y_k, W^{(n)}_{\bm R})\mathrm{d}\mu(\bm R)
	\end{split}
	\end{equation}
	We are interested in the likelihood of the model at iteration $n$, denoted by $\mathcal{L}(W^{(n)})$. Using the expression above, it reads $\mathcal{L}(W^{(n)}) = \mathcal{Q}(W^{(n)}\vert W^{(n)}) + \mathcal{H}(W^{(n)}\vert W^{(n)})$. Using similar arguments as above, the latter quantity can be rewritten the following way
	\begin{equation}
	\begin{split}
	\mathcal{L}(W^{(n)}) &= \underbrace{\frac{1}{K}\sum_{k=1}^K\sum_{j=1}^M P_{jk}\left(\sum_{i=1}^Ny_{ik}\log W_{ij} - W_{ij}\right)}_{\mathcal{Q}(W^{(n)}\vert W^{(n)})}\\
	&-\underbrace{\frac{1}{K}\sum_{k=1}^K\sum_{j=1}^M P_{jk}\log\left(\frac{P_{jk}}{w_j}\right)}_{\mathcal{H}(W^{(n)}\vert W^{(n)})}
	\end{split}
	\end{equation}
	We note that the quantity $-\mathcal{H}(W^{(n)}\vert W^{(n)})$ is exactly the \emph{mutual information} quantity proposed by Loh and Elser \cite[equation 19]{loh2009reconstruction}. 

	Our stopping criterion is now defined as follows. At each iteration, we compute the relative likelihood variation denoted $\nabla \mathcal{L}$. When the value of $\nabla \mathcal{L}$ goes below a given threshold $\eta$, then either the algorithm stops if the current value of $L$ is equal to $L_\text{max}$, or restart with an increased bandlimit $L$ and uses the previously estimated spherical harmonic coefficients as a starting point.

	\subsubsection{Summary}
	The computational time of the algorithm presented in this paper scales directly with the desired resolution of the desired reconstruction. Actually, the most expensive part is the maximization step, which in our case scales as $\mathcal{O}(L^3N_sK)$. It thus shows a clear dependence on the bandlimit $L$, and on the number of pixels on the shell. This motivates the shell-by-shell approach: since inner and outer shells show two disjoints set of parameters (small $L$ and $N_s$ for the inner shell, large $L$ and $N_s$ for the outer shells), it is interesting to gain computation time by parallelization of the shell reconstructions. 

	Also, the increasing bandlimit approach is expected to lead to a better conditioning of the algorithm, limiting the number of iterations and improving the quality of the reconstruction. 

	Finally, we shall point out that the sampling of the rotation group used here is based upon harmonic analysis results on the rotation group. Therefore, one may expect the harmonic approximation of integrals over $\SO{3}$ used in the EMC algorithm to perform equally well compared to the $600$-cell sampling scheme previously introduced \cite{loh2009reconstruction}.


	\section{Spherical shell alignment}\label{section:shellAlignement}
\label{sec:shell}
	Since we have considered each shell separately (which allows an efficient parralelization of the algorithm, and distribution of the data), the shells have to be realigned in order to form a consistent 3D function. Namely, we want to minimize the quadratic error between two successive shells. The problem can be stated as follows, we need to find the rotation $\bm R$ such that
	\begin{equation}\label{eq:position_problem_shell_alignment}
	\underset{\bm R \in \SO{3}}{\arg \min} \: \Vert W_\mathcal{G}^{(s)} - \Lambda(\bm R)W_\mathcal{G}^{(s+1)}\Vert^2_2,
	\end{equation}
	where we have introduced the rotation operator $\Lambda(\bm R)$ defined by $\Lambda(\bm R)f(\bm \Omega) = f(\bm R^T\Omega)$.
	As the features shared by the shell intensities evolve with the radius $q_s$, the spherical shell alignment problem between two consecutive shells is only significant if these are sufficiently close to each other. The alignment procedure is indeed expected to fail in the case where $(q_{s+1}-q_s) > \Delta^\text{SN} q/2$, that is the Shannon-Nyquist theorem is no longer verified in the radial direction. 

	In this paper the shell alignment problem is solved using correlations on the rotation group, using an approach proposed by Kostelec and Rockmore in \cite{kostelec2008ffts}. For an overview concerning the so-called \emph{rotational matching problem}, see for instance \cite{chirikjian2004rotational} and reference therein, especially for the rotational matching problem in biophysics.

	The problem as stated in (\ref{eq:position_problem_shell_alignment}) is equivalent to maximizing the correlation of the two shell intensities,
	\begin{equation}
		\mathrm{Corr}_{s,s+1}(\bm R) = \int_{\mathcal{S}^2} W_\mathcal{G}^{(s)}(\bm \Omega)\overline{\Lambda(\bm R)W_\mathcal{G}^{(s+1)}(\bm \Omega)}\mathrm{d}\bm\Omega
	\end{equation}
 	We remind that the shell intensities are also represented by their estimated spherical harmonic coefficients up to a given bandlimit $L$), respectively $\hat{c}_\ell^{-m}(q_s)$ and $\hat{c}_\ell^{-m}(q_{s+1})$. Using Fourier analysis properties on the rotation group, the correlation above can be expressed the following way
	\begin{equation}\label{eq:correlation_practical}
	\begin{split}
		\mathrm{Corr}_{s,s+1}(\bm R) = \sum_{\ell=0}^{L-1}\sum_{m=-\ell}^\ell\sum_{m'=-\ell}^\ell \hat{c}_\ell^{m}(q_s)\overline{\hat{c}_\ell^{m}(q_{s+1})}\times\\
		\overline{D_\ell^{m,m'}(\bm R)}
	\end{split}
	\end{equation}
	where we have introduced the Wigner-D functions, $D_\ell^{m,m'}$, which are detailed in appendix \ref{appendix_B}. Now, using the same rotation group sampling as in the spherical EMC algorithm, one can evaluate the correlation (\ref{eq:correlation_practical}) efficiently. The rotation $\bm R_j$ that leads to the higher correlation value between the two shells is then used to align the $s+1$ shell with the previous one. Namely, the \emph{aligned} spherical harmonic coefficients read
	\begin{equation}
	\tilde{\hat{c}}_\ell^m(q_{s+1}) = \sum_{n=-\ell}^\ell D_\ell^{m,n}(\bm R_j)\hat{c}_\ell^m(q_{s+1}).
	\end{equation}
	By repetition of the same procedure for succesive shells, we are able to reconstruct efficiently the full 3D intensity function.

	Finally, a few non trivial points are to be mentioned. First, since the correlation is performed numerically on a discrete rotation sample, it is unlikely that the rotation leading to the higher correlation value will be the \emph{exact} one. However, by increasing the size of the rotation sampling, the rotation estimation error will be further reduced. Moreover, from a practical perspective, a good approximation of the correlation can be obtained using only spherical harmonics of low degree, even in the case of large bandlimits $L$ \cite{kostelec2008ffts}. This could be adequatly exploited to obtain low-resolution reconstructions efficiently. 

	Finally one can note that a method as been proposed in \cite{makadia2006rotation} to improve the accuracy of the rotation matching, and could be efficiently applied in our context.

\section{Numerical validation}
\label{sec:numeric}
	In this section, we illustrate our approach with simulations. First, we show how the adaptive spherical EMC can effectively reconstruct the shell intensity for different bandlimits. We illustrate the potential of the approach by considering three typical shells, an inner, middle and outer shell respectively. Theoretical intensity functions corresponding to these three shells are presented in figure \ref{fig:theoreticalEnergyDistribution}.


	The behavior of the algorithm with respect to the number of observations is also studied in detail. 

	Secondly, we investigate the full shell-by-shell reconstruction problem, and analyse the effect of the distance between shells on the resulting resolution. 

	All simulations show results for the Satellite Tobacco Necrosis Virus (STNV, PDB entry: 2BUK), one of the smallest viruses known to date. The reciprocal space size of a pixel is $\Delta q = 0.011$ \AA$^{-1}$. Simulations are performed for a fluence $I_0$ corresponding to $10^{13}$ photons per pulse and $0.1$ $\mu$m$^2$ focal spot area. Corresponding Mean Photon Count (MPC) per pixel are given in the top of figure \ref{fig:theoreticalEnergyDistribution}.

	We note that the number of photons per pulse used here ($10^{13}$) is about ten times higher than those currenlty available at XFEL facilities. This choice is motivated here by the fact that it allows us to use less diffraction patterns to reconstruct the biomolecule,  as the aim of this work is to demonstrate the feasability of the approach. As it will be seen in the next section, using state of the art photon counts will lead to an higher number of diffraction patterns required for a given bandlimit $L$.

	\subsection{Single shell reconstruction}
			\subsubsection{Estimation of the required number of diffraction patterns}
			In this section we address the important issue of the required number of diffraction patterns in single-particle experiments. Namely, we ask the following question: given a specified shell, how many diffraction patterns are needed to reconstruct the intensity defined on this shell, for a given bandlimit $L$ ?

			This question is in general quite complex, and to allow further investigation we need to make some assumptions. The reconstruction of the intensity on the shell is conditional upon the ability to recover the orientation of each diffraction pattern, yet to have sufficient signal-to-noise ratio. It is expected that the number of diffraction patterns required will be governed by a combination of these two factors.

			When the signal-to-noise ratio is high, as expected for large biomolecules and viruses (and already shown by a recent experiment \cite{ekeberg2015three}), the orientation recovery problem drives predominantly the performances of reconstruction algorithms. In the sequel, we perform numerical analysis addressing the question stated previously, taking into account both orientation recovery issue and sufficient signal-to-noise ratio.

			Let us consider a shell of radius $q_s$, and that detector coordinates on this shell are given by equation (\ref{eq:detector_definition_shell}). We note that the number of pixels on the detector $N_s$ scales with the radius $q_s$, given the pixel spacing $\Delta q$. Moreover, we fix a Mean Photon Count per pixel on this shell. Now, we define an equal-area grid on the sphere with HEALPix, with a resolution parameter $\nside$. We draw an uniform random rotation from $\SO{3}$, and apply it to the detector coordinates. Then, we draw $N_s$ samples from the Poisson distribution of parameter $\text{MPC}$. Finally, we analyse which HEALPix pixels are visited by the non-zero pixels of this \emph{rotated detector} and keep trace of this visit. The number of runs of this simulation until each pixel has been visited at least once gives us an estimate of the required number of diffraction patterns on this shell, for a given resolution $\nside$. Repeating this algorithm several times gives a fair estimate of this number. 

			Here, we only analysed the number of required patterns in terms of the resolution parameter $\nside$. Recalling that $\nside$ is linked to the bandlimit $L$ by the approximate sampling theorem (\ref{eq:appendix_sampling_theorem_Healpix}), we link each $\nside$ value with the maximum bandlimit $L$ available for this resolution parameter. Extrapolation between missing $L$-values gives an estimate of the required diffraction patterns for any $L$ value. Note that with this choice of detector geometry and for a given shell, the maximum value of $L$ available is given by 
			$L^{\text{max}} = 1 + \left\lfloor \pi q_s/\Delta q\right\rfloor$. This simply means that the maximum bandlimit available increases with the shell radius.

\begin{figure}
			\includegraphics[width=0.49\textwidth]{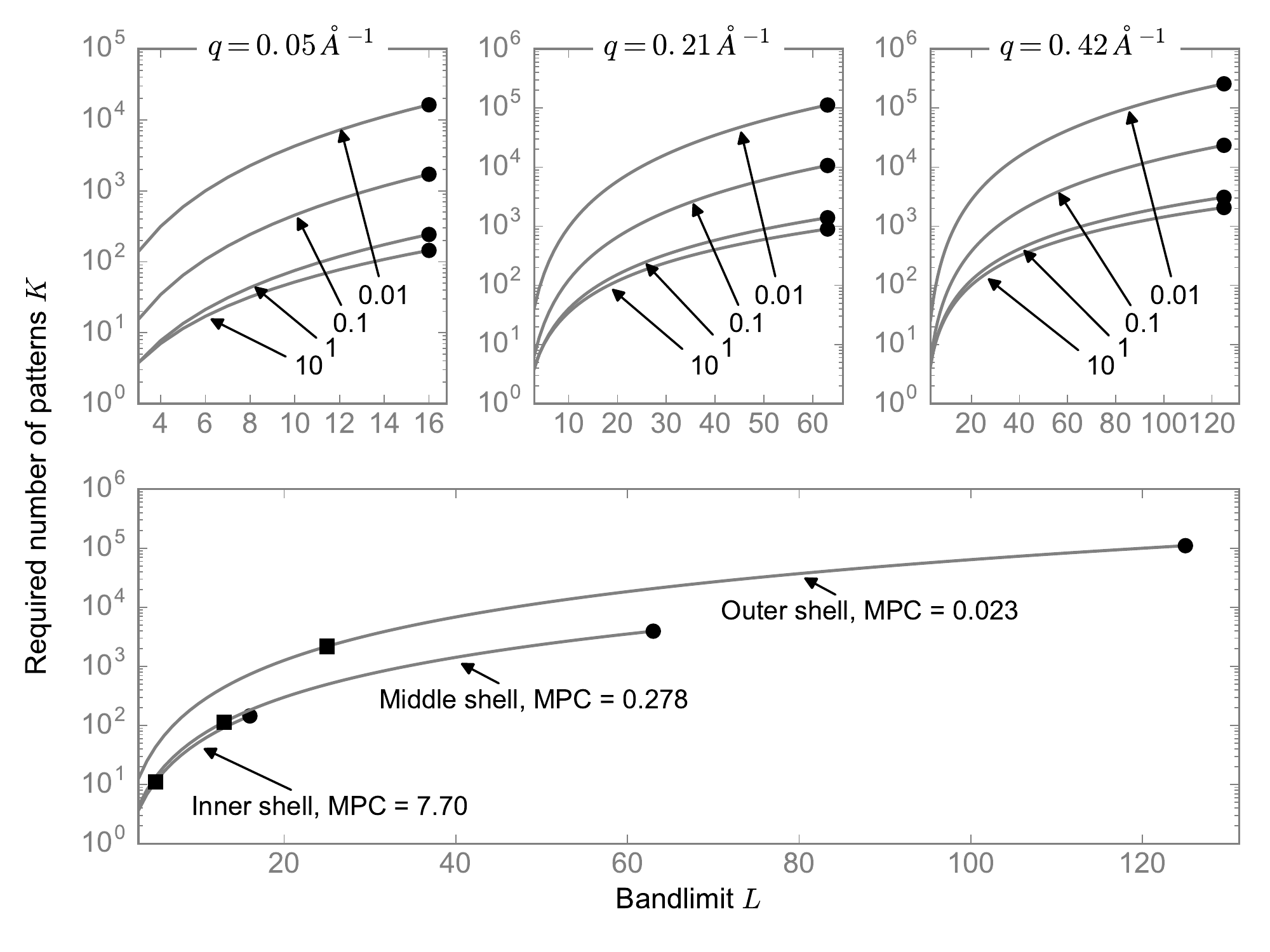}\caption{(Top) Required number of patterns with the bandlimit $L$ and MPC, where each plot correspond to different shell radii. In each plot the following MPC values were used  $\text{MPC} = 10$, $1$, $0.1$, $0.01$. (Bottom) Corresponding required patterns in the case of the STNV virus, using MPC values shown in figure \ref{fig:theoreticalEnergyDistribution}.}\label{fig:requiredPatterns}
			\end{figure}

			The results of the simulation are presented in figure \ref{fig:requiredPatterns}. In the first row, we analysed the number of diffraction patterns needed in the case of three typical shells, inner, middle, and outer, and for different MPC on these shells. The MPC used here are $10$, $1$, $0.1$, $0.01$. First, one sees that for a given shell and bandlimit, the number of required patterns increases as the MPC decreases, as expected. Similarly the patterns required increase with the bandlimit $L$, given a certain shell and MPC.

			For a given bandlimit $L$ and a fixed MPC, the required number of patterns reduces as the shell radius increases. This is the expected behaviour, since at $L$ fixed, the area of each pixel on the sphere increases. Moreover, we note that for each shell, the required number of patterns behaves as a power law of $L$. 

			At the bottom of figure \ref{fig:requiredPatterns}, the number of patterns required in the case of the STNV virus are presented. Note that the number of required patterns increases with the shell radii, due to a lower MPC in the outer shells. Black squares show the theoretical bandlimit of the shell intensity, calculated from figure \ref{fig:theoreticalEnergyDistribution} by thresholding the relative energy per degree below some predefined threshold, typically $10^{-5}$ in our experiments.

		\subsubsection{Multiresolution shell reconstruction}
			In the sequel, we investigate the behavior of the algorithm in three typical cases of shell intensity reconstructions: inner, middle and outer shell cases. The simulations presented here have been performed using 500 diffraction patterns for the inner shell case and 1000 diffraction patterns for the middle and outer shell cases. We make use of the adaptive shell EMC algorithm presented earlier. This algorithm provides an adaptive reconstruction method based on the spherical harmonic decomposition of the shell intensity up to a maximum bandlimit $L_\text{max}$.

\begin{figure*}
			\includegraphics[width=\textwidth]{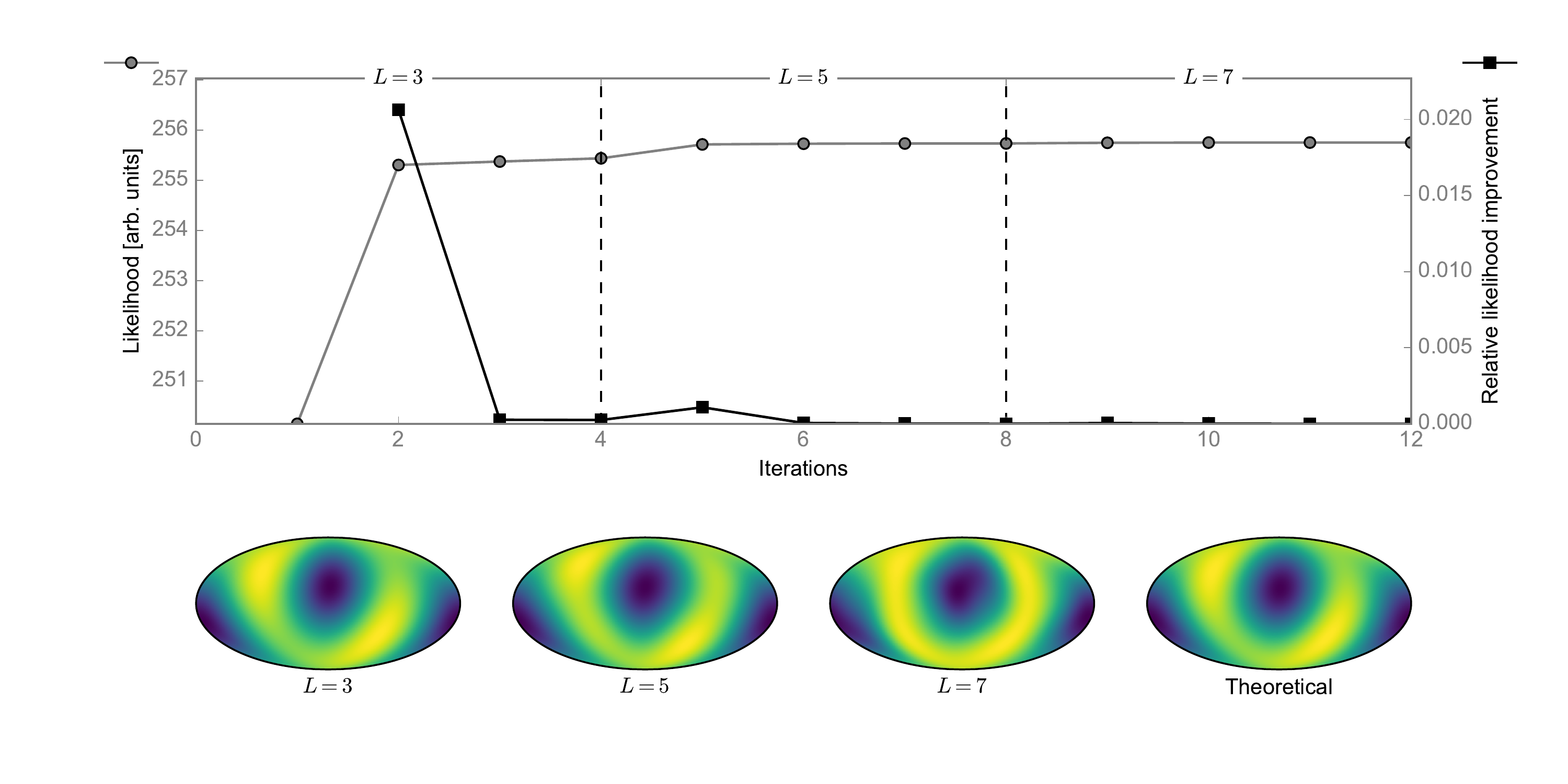}\caption{Adaptive intensity reconstruction of the inner shell case ($q = 0.05 $ \AA$^{-1}$). Top: Evolution of the likelihood with the iteration number and corresponding relative likelihood improvements. Vertical lines show changes in the bandlimit, respectively $L=3, 5, 7$. Bottom: Corresponding reconstructions at the end of each fixed bandlimit and comparison with the theoretical intensity. Colormaps were adapted to enhance the contrast of the vizualization.}\label{fig:reconstructionInner}
			\end{figure*}

			The inner shell case is represented in figure \ref{fig:reconstructionInner}. From figure \ref{fig:theoreticalEnergyDistribution}, the expected theoretical bandlimit is $\Llim = 7$. One can notice that the likelihood improvements become rapidly very small, and that the number of iterations for each bandlimit $L$ is given by the minimum number of iterations imposed (in our case 4). This behavior is easily seen from the low-bandlimit distribution of the inner shell. As seen in the reconstruction provided, the main features of the shell intensity are already reconstructed at $L = 3$, as expected.

\begin{figure*}
			\includegraphics[width=\textwidth]{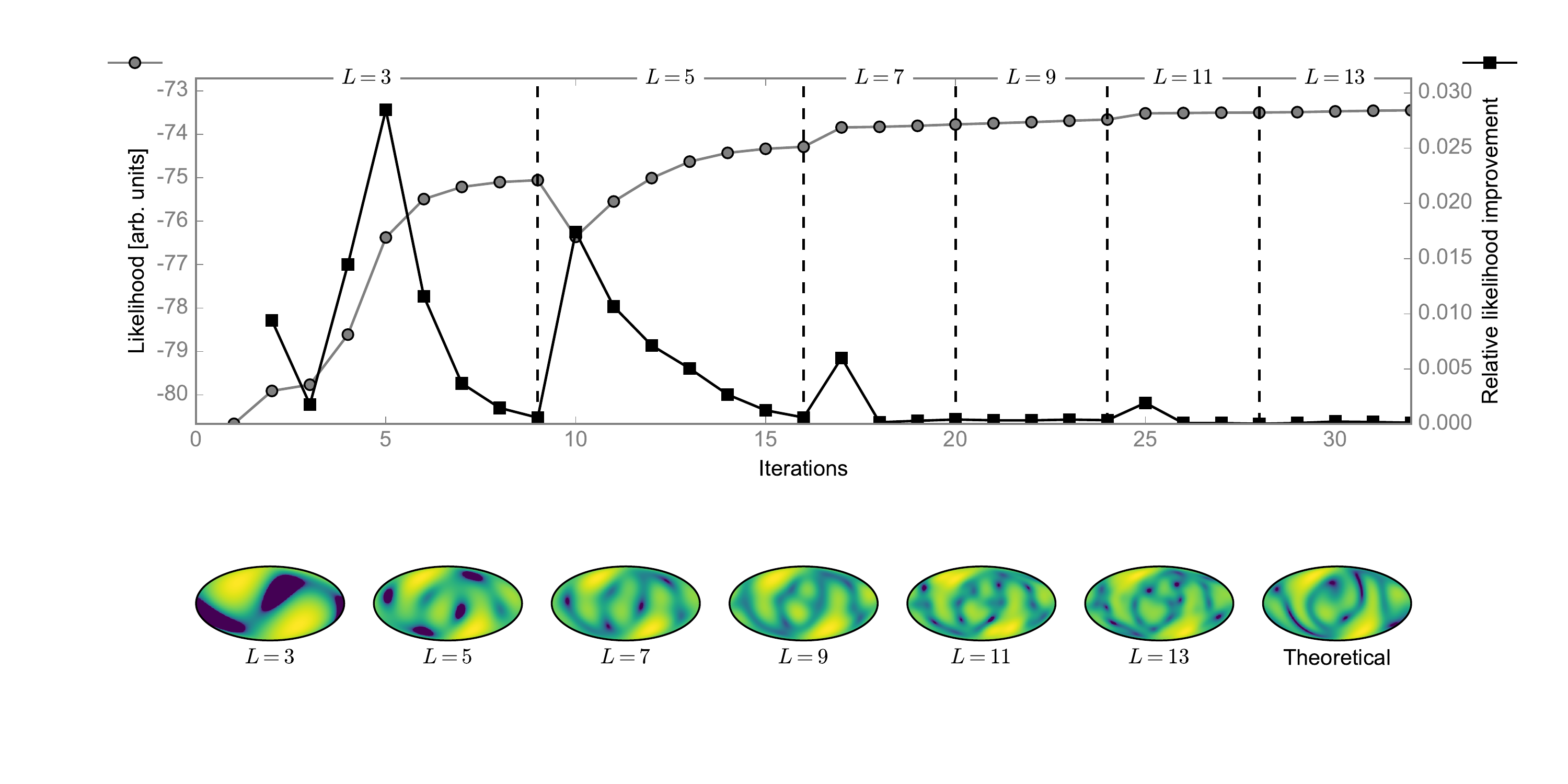}\caption{Adaptive intensity reconstruction of the middle shell case ($q = 0.21 $ \AA$^{-1}$). Top: Evolution of the likelihood with the iteration number and corresponding relative likelihood improvements. Vertical lines show changes in the bandlimit, respectively $L=3, 5, 7, 9, 11, 13$. Bottom: Corresponding reconstructions at the end of each fixed bandlimit and comparison with the theoretical intensity. Colormaps were adapted to enhance the contrast of the vizualization.}\label{fig:reconstructionMid}
			\end{figure*}

			The middle shell case is represented in figure \ref{fig:reconstructionMid}. Again, using figure \ref{fig:theoreticalEnergyDistribution}, we see that the expected theoretical bandlimit is now $\Llim = 13$. The likelihood is improving over the different bandlimits, tending to smaller increments as the bandlimit increases. The reconstructions provided shows how the accumulation of spherical harmonic coefficients of higher degree improves the accuracy of the reconstruction. One can note that low-bandlimit reconstructions exhibit negative values (up to $L = 7$, included); therefore the negative values are thresholded to some small constant (here $10^{-4}$). This stems from the fact that truncation of a spherical harmonic expansion of the non-negative function does not preserve the non-negativity of the function.  

\begin{figure*}
			\includegraphics[width=\textwidth]{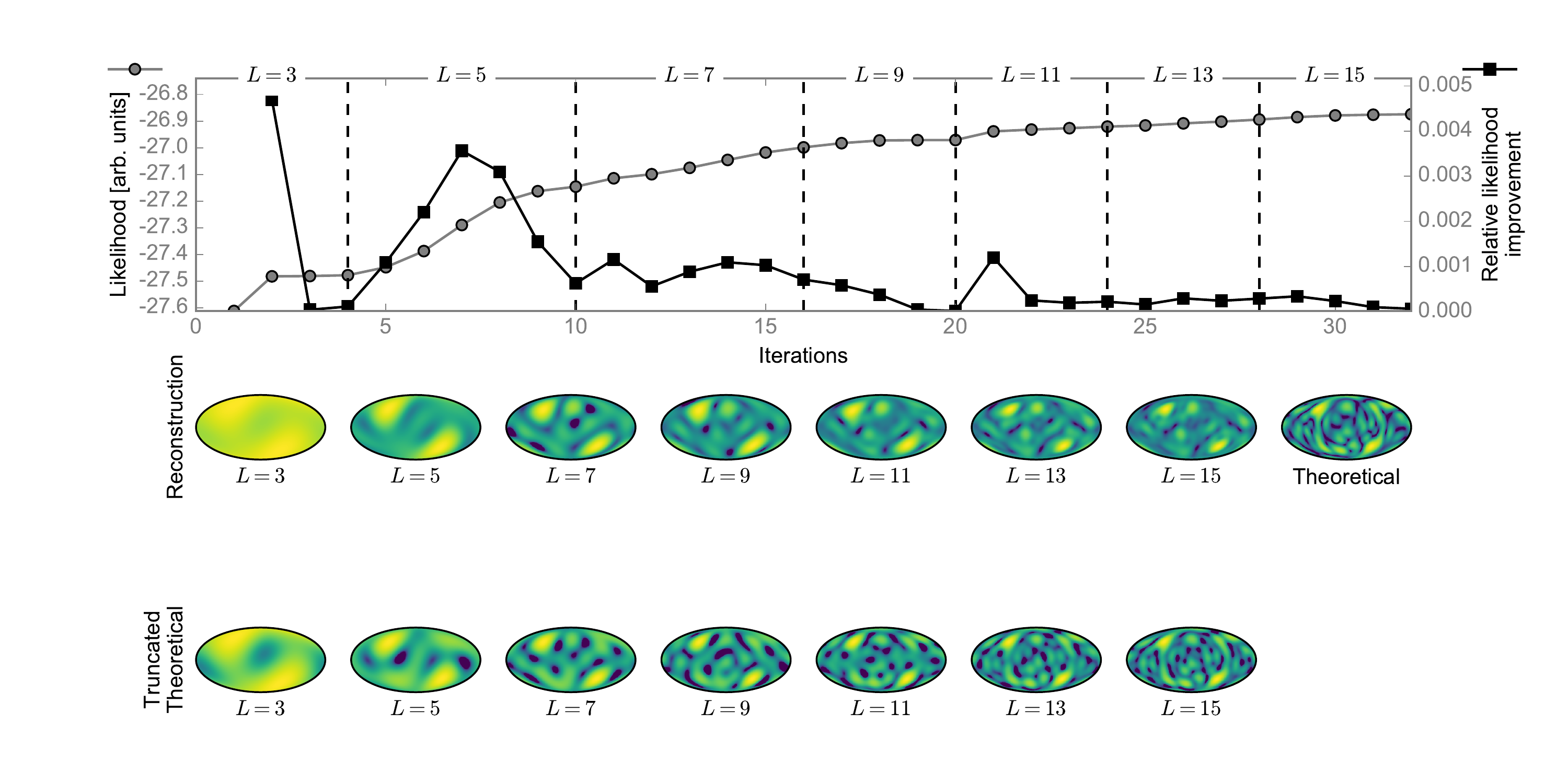}\caption{Adaptive intensity reconstruction of the outer shell case ($q = 0.42 $ \AA$^{-1}$). Top: Evolution of the likelihood with the iteration number and corresponding relative likelihood improvements. Vertical lines show changes in the bandlimit, respectively  $L=3, 5, 7, 9, 11, 13, 15$. Bottom: Corresponding reconstructions at the end of each fixed bandlimit and comparison with respective truncated theoretical intensities.}\label{fig:reconstructionOuter}
			\end{figure*}

			Finally, the outer shell case is represented in figure \ref{fig:reconstructionOuter}. In this case, the theoretical bandlimit is $\Llim = 25$, as seen in figure \ref{fig:theoreticalEnergyDistribution}. Here, the likelihood tends to improve constantly as the bandlimit increases before stabilizing from $L = 13$. Reconstructions provided show the frequency-like improvement as the bandlimit $L$ increases. Truncated theoretical intensities for the respective $L$ values are also provided. However, one can notice the difference between the theoretical and reconstructed intensities. The reasons are twofold: first when estimating low-$L$ intensity functions, there is aliasing of power of higher degree coefficients ($\ell > L$), resulting in an higher overall energy in the reconstruction than in the truncated theoretical intensity. This is well shown by results in the cases $L=3, 5$ and $L=7$. The second results from an insufficient number of diffraction patterns, since for $L > 7$ the reconstruction barely improves, whereas the truncated theoretical intensity continues to converge to the theoretical intensity.

	\subsection{Full shell-by-shell reconstruction}
		
		\begin{figure*}
		\centering
		\includegraphics[width=\textwidth]{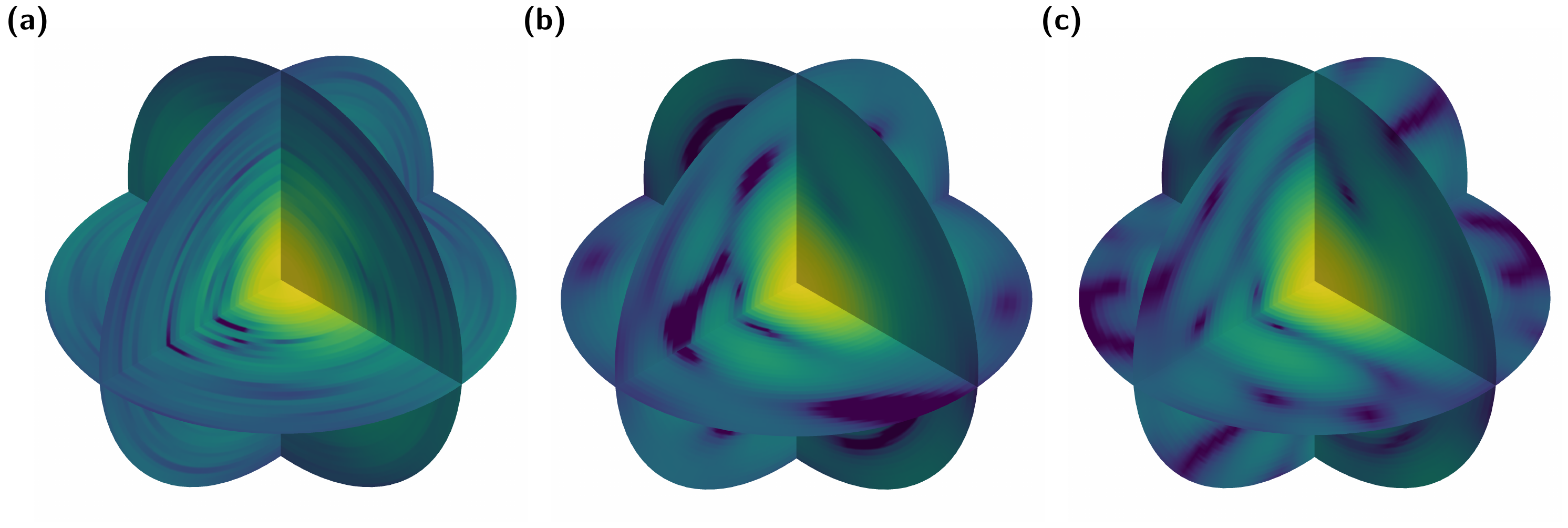}\caption{Full reconstruction of the intensity function for the STNV virus. \textbf{\textsf{(a)}} Reconstruction from 1000 diffraction patterns using 40 shells spaced by $\Delta q$. The reconstruction is bandlimited at $L = 7$. \textbf{\textsf{(b)}} Corresponding theoretical intensity function bandlimited at $L = 7$. \textbf{\textsf{(c)}} Theoretical intensity function.}\label{figure:fullReconstruction}
		\end{figure*}
		\begin{table}
			\begin{tabular}{*3c}
			\multicolumn{3}{l}{$R$-factor values}\\
			\cmidrule[1pt]{1-3}
			&Truncated &Theoretical\\
			\cmidrule[0.3pt]{1-3}
			$L = 3$& 0.152 & 0.171\\
			$L = 5$& 0.143 & 0.148 \\
			$L = 7$& 0.0979 & 0.0998\\
			\cmidrule[1pt]{1-3}
			\end{tabular}
			\caption{$R$-factor values computed according to (\ref{eq:defRFactor}) with respect to the $L$-truncated theoretical intensity or the theoretical intensity, for different $L$ values.}\label{table:Rfactors}
		\end{table}

		The shell-by-shell approach offers an efficient and faster method to distribute computation of the reconstructions. However, by considering the shells independently the radial information is lost. This radial information has to be retrieved during an alignment process, as described in section \ref{section:shellAlignement}.

		Using 1000 diffraction patterns, we reconstructed each shell intensity independently up to bandlimit $L = 7$ and maximum frequency $q = 0.42$ \AA$^{-1}$, for a reciprocal space spacing $\Delta q = 0.01$ \AA$^{-1}$ between the shells. The shells were then realigned using the above mentioned correlation method. Results are presented in figure \ref{figure:fullReconstruction}.

		To evaluate numerically the performance of the method, we introduce an error metric based on the well-know $R$-factor, which is defined such that
		\begin{equation}\label{eq:defRFactor}
			R = \frac{\sum_p \vert I_r(\bm q_p) - I_{\text{th}}(\bm q_p)\vert }{\sum_p I_r(\bm q_p)}
		\end{equation}
		where $I_r$ denotes the reconstructed intensity, and $I_{\text{th}}$ is the theoretical intensity. Here $\bm q_p$ denotes the 3D-spherical grid points. To illustrate the reconstruction performances with the bandlimit $L$, we have computed the $R$-factor for $L=3, 5, 7$ reconstructions. We distinguished two cases, one where the theoretical intensity is truncated up to degree $L$, and the other where the full theoretical intensity is considered. As shown by the results in \ref{table:Rfactors}, the $R$-factor decreases as $L$ increases, meaning that the reconstructed intensity function is improved over $L$. Also one notices that the $R$-factors for the truncated theoretical intensity is smaller than for the full theoretical intensity, which is consistent with the fact that the full intensity function is more \emph{complex} than the truncated one.

		Finally, one can note that different sources of estimation error arise in the shell-by-shell approach. One is related to the shell intensity estimation problem, where the finite number of diffraction patterns limits the actual accuracy of the reconstruction. The second one is due to the radial decoupling between the shells, which leads to an realignment error. The latter can be minimized by either increasing the size of the grid on which the $SO(3)$ correlation is performed, or either by using refinment methods such as in \cite{makadia2006rotation}. As a consequence, the algorithm presented here may not be as accurate as the original EMC algorithm in the case of low SNR datasets. It is actually expected that the proposed approach is a compromise improved computational performance of a shell-by-shell approach and noise tolerance.

\section{Conclusion}
Here we have shown that a spherical harmonic basis is a convenient way to design an incremental, parallelizable approach to the reconstruction of 3D intensity functions from 2D XFEL diffraction patterns. The proposed shell-by-shell algorithm allows control over radial and angular resolution, which can be gradually increased during the reconstruction. As a shell-by-shell approach is intrinsically parallelizable, it is potentially an efficient way to analyse the very large datasets expected in experiment. Using numerical simulations, we have studied how the number of patterns required for convergence depends on the signal and resolution. 

\acknowledgments{Nicolas Le Bihan's research was supported by the ERA, European Union, through the International Outgoing Fellowship (IOF GeoSToSip 326176) program of the 7th PCRD. AVM was supported by the Australian Research Council's DECRA funding scheme (DE140100624) and Centre of Excellence programmes.}


\appendix
	\section{Fourier analysis on the sphere}\label{appendix_A}
	\subsection{Spherical harmonics}
		The spherical harmonic functions, or in short spherical harmonics, extend Fourier analysis to the 2-sphere $\mathcal{S}^2$. Formally, the spherical harmonic of degree $\ell$ and order $m$ is given by
		\begin{equation}
		Y_\ell^m(\theta,\phi) = \sqrt{\frac{(2\ell+1)}{4\pi}\frac{(\ell-m)!}{(\ell+m)!}}P_\ell^m(\cos \theta)\exp(\mathrm{i}\phi),
		\end{equation}
		where $P_\ell^m$ is the associated Legendre polynomial (or generalized Jacobi polynomial). For each integer degree $\ell \geq 0$ corresponds $2\ell +1$ orders $m$, such that $\vert m \vert \leq \ell$. The spherical harmonic functions form a complete orthonormal basis on the 2-sphere $\mathcal{S}^2$ for square integrable functions. As a consequence any square-integrable complex-valued function $f: \mathcal{S}^2 \rightarrow \mathbb{C}$, $f\in L^2(\mathcal{S}^2)$ can be decomposed onto spherical harmonics,
		\begin{equation}\label{eq:appendix_infinite_SH_decomposition}
		f(\theta,\phi) = \sum_{\ell = 0}^\infty\sum_{m=-\ell}^\ell f_\ell^m Y_\ell^m(\theta,\phi)
		\end{equation}
		where we have introduced the spherical harmonic coefficients $f_\ell^m$, which are given by projection onto the respective spherical harmonic $Y_\ell^m$ via the canonical inner product on $L^2(\mathcal{S}^2)$, 
		\begin{equation}
		f_\ell^m \defeq \ip{f,Y_\ell^m}_{\mathcal{S}^2} = \int_{\mathcal{S}^2}f(\theta,\phi)\overline{Y_\ell^m(\theta,\phi)}\sin\theta \:\mathrm{d}\theta\mathrm{d}\phi.
		\end{equation}
		These coefficients are complex-valued, and the properties of the function $f$ (real-valued, symmetries) are encoded in the coefficients. As in classical Fourier analysis, spherical harmonics exhibit Parseval relation, that is the energy of $f$ is preserved in the spherical harmonic domain,
		\begin{equation}
			\Vert f \Vert^2 \defeq \int_{\mathcal{S}^2} \vert f(\theta,\phi)\vert^2\sin\theta \:\mathrm{d}\theta\mathrm{d}\phi = \sum_{\ell=0}^\infty\sum_{m=-\ell}^\ell \vert f_\ell^m\vert^2.
		\end{equation}
		It is often convenient to introduce a related rotation-invariant quantity, known as the energy per degree $E_\ell$, defined by $E_\ell = \sum_{m=-\ell}^\ell \vert f_\ell^m\vert^2$.

		For numerical purposes the infinite series expansion in the spherical harmonic decomposition in (\ref{eq:appendix_infinite_SH_decomposition}) is not desirable. Rather, we consider bandlimited functions on the sphere. We say that a function $f$ is bandlimited at $L$ (or equivalently $L$-bandlimited) if for all $\ell \geq L$, the spherical harmonic coefficients are identically zero $f_\ell^m = 0$. In the following, we consider $L$-bandlimited functions only.

	\subsection{Spherical harmonic transforms with HEALPix}
		
		The forward and inverse Spherical Harmonic Transform (SHT) are given by 
		\begin{align}
		f_\ell^m &= \int_{\mathcal{S}^2} f(\theta,\phi) \overline{Y_\ell^m(\theta,\phi)}\sin\theta\mathrm{d}\theta\mathrm{d}\phi&\text{(SHT)},\label{eq:appendix_SH_forward}\\
		f(\phi,\theta) &= \sum_{\ell=0}^{L-1}\sum_{m=-\ell}^\ell f_\ell^mY_\ell^m(\phi,\theta)&\text{(inv.SHT)}.
		\end{align}
		The computation of the forward SHT requires the evaluation of an integral over the 2-sphere $\mathcal{S}^2$ for each coefficient $f_\ell^m$. The evaluation of such integrals can be done by conveniently sampling the 2-sphere, \ie distributing nodes on the surface on the sphere in order to obtain a quadrature formula. To this aim, a certain number of sampling schemes have been already proposed \cite{mcewen2011novel,driscoll1994computing,doroshkevich2005gauss,gorski2005healpix,crittenden1998exactly}, often motivated by the analysis of the Cosmic Microwave Background. In this work, we use the HEALPix sampling scheme \cite{gorski2005healpix}, which has been designed for high performance, fast and accurate computation of spherical harmonics on the sphere. The sphere is tessellated into curvilinear equal-area pixels, where the pixel centers are distributed on lines of constant latitude allowing faster computation of spherical harmonic functions due to the separation of angular variables in the spherical harmonics. The HEALPix sampling scheme is hierarchical and provides different levels of resolution through a parameter $\nside$. The number of pixels $\npix$ at resolution $\nside$ is given by
		\begin{equation}\label{eq:appendix_sampling_theorem_Healpix}
		\npix = 12\times \nside^2,
		\end{equation}
		where the parameter $\nside$ is a power of $2$, due to the construction of the grid.\footnote{This fact arises from the hierarchical decomposition of the grid, where the next resolution level is obtained by dividing each pixel into four equal area pixels.}

		Although HEALPix grid do not exhibit an exact sampling scheme, a very accurate estimate of the spherical harmonics coefficients in (\ref{eq:appendix_SH_forward}) is obtained if the following condition is fulfilled;
		\begin{equation}
			L \leq 2\nside+1.
		\end{equation}
		This last formula stands for an approximate sampling scheme on the grid. 

\section{Rotation group sampling}\label{appendix_B}

	\subsection{Definition}
	The special orthogonal group in three dimensions $\SO{3}$ denotes the set of real matrices $\bm R \in \mathbb{R}^{3\times 3}$ which satisfy
	\begin{equation}\label{so3def}
	\bm R\bm R^T = \bm I_3 \quad \text{and} \quad \det \bm R = 1,
	\end{equation}
	where $\bm R^T$ is the transpose of the matrix $\bm R$, $\det \bm R$ its determinant and $\bm I_3$ the $3\times 3$ identity matrix \cite{altmann2005rotations}. It is commonly parametrized by Euler angles, $(\alpha,\beta,\gamma)$ where $\alpha,\gamma \in [0,2\pi)$ and $\beta \in [0,\pi)$. In the $zyz$ convention, the matrix $\bm R(\alpha,\beta,\gamma)$ reads
	\begin{equation}
	\bm R(\alpha, \beta, \gamma) = \bm R_z(\alpha)\bm R_y(\beta) \bm R_z(\gamma),
	\end{equation}
	where $\bm R_z, \bm R_y$ denote rotations around canonical axes $z$ and $y$, respectively. These rotations reads explicitely
	\begin{align}\label{eulerformsmatrixz}
	\bm R_z(\alpha) &= \begin{pmatrix}
	\cos \alpha & - \sin \alpha&0\\
	\sin\alpha & \cos \alpha&0\\
	0&0&1\\
	\end{pmatrix},\\\label{eulerformsmatrixy}
	\bm R_y(\beta)&= \begin{pmatrix}
	\cos \beta & 0&\sin\beta\\
	0&1&0\\
	-\sin\beta & 0&\cos \beta\\
	\end{pmatrix}.
	\end{align}

	\subsection{Harmonic analysis on the rotation group}
	Harmonic analysis on the rotation group $\SO{3}$ is conveyed by the irreducible representations of $\SO{3}$. This is direct consequence of the Peter-Weyl theorem in the $\SO{3}$ case \cite{barut1986theory}. Namely, any square-integrable function $f: \SO{3}\rightarrow \mathbb{C}$ can be decomposed as
	\begin{equation}
		f(\bm R)  = \sum_{\ell = 0}^\infty\sum_{m,n = -\ell}^\ell f_\ell^{m,n}D_\ell^{m,n}(\bm R)
	\end{equation}
	where the $D_\ell^{m,n}$ functions are the Wigner-D functions and $f_\ell^{m,n}$ are the Fourier coefficients obtained following
	\begin{equation}\label{eq:fourier_transform_SO3}
	f_\ell^{m,n} = \int_{\SO{3}}f(\bm R)\overline{D_\ell^{m,n}(\bm R)}\mathrm{d}\mu(\bm R).
	\end{equation}
	The Wigner-D functions are conveniently expressed using the $zyz$-Euler angles parametrization, that is 
	\begin{equation}
		D_\ell^{m,n}(\alpha,\beta,\gamma) = e^{-\mathrm{i}\alpha}d_\ell^{m,n}(\beta)e^{-\mathrm{i}\gamma},
	\end{equation}
	where $d_\ell^{m,n}(\beta)$ is a polynomial in $\cos(\beta/2)$ and $\sin(\beta/2)$.

	The evaluation of the Fourier transform (\ref{eq:fourier_transform_SO3}) can be done using Fast Fourier Transforms on the rotation group, as first developed in \cite{kostelec2008ffts}.
	
	\subsection{Sampling theorem on the rotation group}
	Let us consider a $L$-bandlimited function $f$ on the rotation group, that is for all $\ell \geq L$, $f_\ell^{m,n} = 0$. It is possible to compute its Fourier transform exactly using the following equiangular sampling
	\begin{equation} 
	\alpha_{j_1} = \frac{2\pi j_1}{2L}, \: \beta_{j_2} = \frac{\pi}{4L}\left(2j_2+1\right), \: \gamma_{j_3} = \frac{2\pi j_3}{2L}
	\end{equation}
	where the indices $j_1,j_2,j_3 \in \left\lbrace 0,1,\ldots, 2L-1\right\rbrace$. Equation (\ref{eq:fourier_transform_SO3}) then reads
	\begin{equation}
	f_\ell^{m,n} = \sum_{j}w_{j}f(\bm R_j)\overline{D_\ell^{m,n}(\bm R_j)}
	\end{equation}
	where we have introduced a single index $j = (j_1,j_2,j_3)$. The weights $w_j = w_{j_1,j_2,j_3}$ satisfy $\sum_{j} w_{j} = 1$ and read
	\begin{equation}
	 w_{j_1,j_2,j_3} = \frac{1}{4L^2}\left(\sum_{k=0}^{L-1}\frac{1}{2k+1}\sin\left[\beta_{j_2}(2k+1)\right]\right) \sin \beta_{j_2}.
	\end{equation}

	This sampling of the rotation group is used throughout this work, at the heart of the EMC algorithm and during the shell alignment process. 

	Note that in the EMC algorithm, the above sampling set is used to approximate the value of the integral of some function. This operation corresponds actually to computing the first Fourier coefficient $f_0^{0,0}$. It can be shown that using only half the samples in each angle in the sampling described above is sufficient (aliasing in the higher order coefficients is therefore tolerated). This remark allows us to obtain $L^3$ samples instead of $8L^3$, however this was not implemented here since the oversampling of the rotation group allows a better approximation of the intermediate quantity $\mathcal{Q}(W\vert W^{(n)})$.



%

\end{document}